# Generation of Anisotropic Massless Dirac Fermions and Asymmetric Klein Tunneling in Few-Layer Black Phosphorus Superlattices

Zhenglu Li, Ting Cao, Meng Wu, and Steven G. Louie[*]

*Department of Physics, University of California at Berkeley, Berkeley, California 94720, USA and Materials Sciences Division, Lawrence Berkeley National Laboratory, Berkeley, California 94720, USA*

[*]E-mail: sglouie@berkeley.edu

**Abstract**

Artificial lattices have been employed in many two-dimensional systems, including those of electrons, atoms and photons, in a quest for massless Dirac particles with flexibility and controllability. Periodically patterned molecule assembly and electrostatic gating as well as moiré pattern induced by substrate, have produced electronic states with linear dispersions from isotropic two-dimensional electron gas (2DEG). Here we demonstrate that massless Dirac fermions with tunable anisotropic characteristics can, in general, be generated in highly anisotropic 2DEG under slowly varying external periodic potentials. For patterned few-layer black phosphorus superlattices, the new chiral quasiparticles exist exclusively in an isolated energy window and inherit the strong anisotropic properties of pristine black phosphorus. These states exhibit asymmetric Klein tunneling with the direction of incidence for wave packet with perfect transmission deviating from normal incidence by more than $50°$ under an appropriate barrier orientation.

**Main text**

The unusual relativistic-like quasiparticles in graphene, known as massless Dirac fermions, intrinsically come from the symmetry of its honeycomb crystal structure [1,2] and therefore can be designed and manipulated in other systems as first theoretically predicted [3,4]. Many experimental efforts have been successfully performed to generate Dirac quasiparticles – in common isotropic two-dimensional electron gas (2DEG) on metal or in semiconductor quantum wells under slowly varying periodic potentials formed by molecule assembly or patterned gate [5-10], as well as with ultra-cold atoms trapped in honeycomb optical crystal structures [11,12] and photons confined in photonic crystals [13,14]. Similarly, lattice mismatch between graphene and substrate can introduce a long-range moiré superlattice potential, giving rise to new generation of Dirac points in graphene [15-22]. These platforms combine advantages of an extended degree of control and various foreign modifications [8], making much physics accessible such as the interplay between Dirac fermions with strong correlation [6] and spin-orbit coupling [9], and Hofstadter butterfly effect [17-21]. Simple symmetry argument prescribes a trigonally modulating potential to generate massless Dirac fermions from an isotropic system [3,8]. However, previous studies do not address the questions: 1) whether these novel quasiparticles can arise from highly anisotropic host states, and 2) what new physical insights and phenomena one can obtain from such systems?

In this work, we show that massless Dirac fermions with chiral character can in general be generated from anisotropic 2DEG systems. We moreover propose that semiconducting few-layer black phosphorus [23,24] is an ideal system to achieve this aim upon electron or hole doping [25,26]. Electron and hole carriers in black phosphorus exhibit strong anisotropy with their effective masses along the two orthogonal crystal axes ($m_x^*$ and $m_y^*$) differing by an order of magnitude [27-29]. With properly designed periodic potential (that can be realized with patterned molecule assembly or electrostatic gating under laboratory conditions), we predict that highly anisotropic massless Dirac fermions are generated and that the ratio of the group velocities along the two crystalline directions reaches $(m_x^*/m_y^*)^{\frac{1}{2}} \sim 1/3$. These anisotropic quasiparticles moreover exist within an isolated energy window separating from other states, hence are expected to be quite measurable and controllable.

Unlike the Klein tunneling process in graphene with carriers normal incident upon a potential barrier always experiencing perfect transmission [30,31], we find that the anisotropic massless Dirac fermions generated in black phosphorus superlattices allow for an asymmetric Klein tunneling so that the directions of normal incidence and perfect transmission can be noncollinear and differed by more than 50°. The anisotropic massless Dirac fermions and the asymmetric Klein tunneling, which may be switched and tuned, provide new degrees of freedom in designing potential devices based on chiral massless fermions.

Monolayer and few-layer black phosphorus is a direct-gap semiconductor with a band gap ~ 1 eV at the Brillouin zone center. We shall use this case as a generic example of an anisotropic massive 2DEG. Let us begin with the effective Hamiltonian for the Γ valley of few-layer black phosphorus [29,32],

$$H_{2DEG}(\boldsymbol{p}) = \frac{p_x^2}{2m_x^*} + \frac{p_y^2}{2m_y^*}, \tag{1}$$

where $p_x$ and $p_y$ are the crystal momenta along the *x* and *y* direction, respectively. Eq. (1) is valid around both the conduction band minimum and valence band maximum, where the band disperses much faster along the *x* direction (armchair direction) than along the *y* direction (zigzag direction) [Figs. 1(a)–(b)]. An external periodic potential varying much slower than the interatomic distances would create a superlattice structure, mixing the states in the Γ valley. For an isotropic 2DEG, with $m_x^* = m_y^*$, a triangular potential would strongly mix states near three $\boldsymbol{K}_j$ ($j$ = 1, 2, 3) points (with $\boldsymbol{K}_j = \boldsymbol{p}_j/\hbar$, determined by the periodicity and orientation of the external potential) with degenerate energy, leading to Dirac fermions [3,9]. In an anisotropic 2DEG, $m_x^* \neq m_y^*$ (for electron-doped few-layer black phosphorus, we take $m_x^* = 0.15 m_e$ and $m_y^* = 1.18 m_e$, where $m_e$ is the free electron mass [27]); however, in general, one can still choose three $\boldsymbol{K}_j$ points (with degenerate energy) and mix these states with some scattering strength from a designed potential, i.e., solving an inverse problem.

As a demonstration, we pick up the following three $\boldsymbol{K}_j$ points,

$$\boldsymbol{K}_1 = (K_0, 0), \quad \boldsymbol{K}_2 = \left(-\frac{K_0}{2}, \frac{\sqrt{3}}{2}\sqrt{\frac{m_y^*}{m_x^*}} K_0\right), \quad \boldsymbol{K}_3 = \left(-\frac{K_0}{2}, -\frac{\sqrt{3}}{2}\sqrt{\frac{m_y^*}{m_x^*}} K_0\right), \tag{2}$$

where $K_0$ is a free parameter scaling the periodicity of the external potential $U(\boldsymbol{r})$. (For concrete illustration, we shall set the periodicity of $U(\boldsymbol{r})$ as 2.35 nm in plotting the figures in the main text.) In Eq. (2), we demands the $\boldsymbol{K}_j$ points satisfy the condition [Figs. 1(c)–(d)]: $E(\hbar \boldsymbol{K}_1) = E(\hbar \boldsymbol{K}_2) = E(\hbar \boldsymbol{K}_3)$ where $E(\hbar \boldsymbol{K}_j)$ is the energy eigenvalue of $H_{2DEG}(\hbar \boldsymbol{K}_j)$. With the degeneracy condition satisfied, the choice of the three $\boldsymbol{K}_j$ points still has many possibilities and may lead to rich phases described by a generalized Weyl equation [33] (in general tilted Dirac cones or open Fermi surfaces), and even higher pseudospin fermions can be generated with more $K_j$ included [34,35]. The external potential would lead to mixing of states near those three $K_j$ points, then having reciprocal lattice vectors given by $G_1 = K_2 - K_1$ and $G_2 = K_3 - K_1$, while $G_3 = G_1 - G_2$ is a dependent vector [Fig. 1(d)].

As a first example, we consider a simple external potential of the following form [3,9],

$$U(r) = 2W[\cos(G_1 \cdot r) + \cos(G_2 \cdot r) + \cos(G_3 \cdot r)], \qquad (3)$$

where $W$ defines the strength of the external potential. The real-space distribution of this potential is plotted in Fig. 2(a), with $W = 0.1$ eV. We define $k$ as a small wavevector, i.e. $|k| \ll |K_j|$, and expand the superlattice wavefunction near the three $K_j$ points as a linear combination of $|j\rangle = e^{i(K_j+k)\cdot r}$. In this basis, the Hamiltonian up to the first order in $k$ is given by $H'_{2DEG} = H_0 + H_1$, with $\langle j|H_0|j'\rangle = W(1 - \delta_{jj'})$ and $\langle j|H_1|j'\rangle = \hbar v_j \cdot k \delta_{jj'}$, where $\delta_{jj'}$ is the Kronecker delta function and $v_j = (v_{jx}, v_{jy}) = \left(\frac{\hbar K_{jx}}{m_x^*}, \frac{\hbar K_{jy}}{m_y^*}\right)$. The eigenvalues of the states for $k = 0$ (Hamiltonian $H_0$) are $-W, -W, 2W$. The resulting two degenerate states are $\frac{1}{\sqrt{2}}(0,1,-1)^T$ and $\frac{1}{\sqrt{6}}(2,-1,-1)^T$, and we may construct a 2×2 subspace with them [3,9]. The Hamiltonian at finite $k$ in the new basis reads $H' = \frac{\hbar v_0}{2}(-\sigma_z k_x - \gamma_0 \sigma_x k_y)$ where $v_0 = \frac{\hbar K_0}{m_x^*}$, $\gamma_0 = \sqrt{\frac{m_x^*}{m_y^*}}$ and $\sigma_{x,y,z}$ are Pauli matrices. With a unitary transform [3,9], we arrive at

$$H(k) = \frac{\hbar v_0}{2}(\sigma_x k_x + \gamma_0 \sigma_y k_y). \qquad (4)$$

The Hamiltonian $H$ resembles the low-energy effective Hamiltonian in graphene [1,2], except that an extra factor $\gamma_0$ renormalizes the second term. The energy eigenvalues and the corresponding eigenvectors are therefore given by

$$E(k) = \frac{\lambda \hbar v_0}{2}\sqrt{k_x^2 + \gamma_0^2 k_y^2}, \qquad (5)$$

$$\psi_\lambda(k) = \begin{pmatrix} 1 \\ \lambda e^{i\phi_s} \end{pmatrix}, \qquad (6)$$

where $\lambda = \pm 1$ denoting the upper and lower bands, and $\phi_s$ is the polar angle of the pseudospin vector associated with $\psi_\lambda(k)$. Here we use the notation "s" for pseudospin. The energy dispersion is linear and these quasiparticles behave as massless Dirac fermions under the modulation of the periodic external potential. There exists two inequivalent Dirac points at $K = K_1$ and $K' = -K$. The group velocity of the linear dispersion varies with the direction of $k$, resulting in an anisotropic Dirac cone band structure. The anisotropy in the group velocity, $\frac{|v_y|}{|v_x|} = \gamma_0$, originates from the anisotropic effective masses intrinsic to the original anisotropic 2DEG. If $K_1 + K_2 + K_3 = 0$ [Eq. (2)], then it can be shown that electron-hole symmetry exists for this model potential.

The above analytic results are for the potential given in Eq. (3) from a perturbative analysis for small $k$, and we have also numerically (see Supplemental Material [36]) solved for the band structure in the whole superlattice Brillouin zone [Figs. 2(b)–(c)], and find that the Dirac cone indeed disperses much faster along the $k_x$ direction than along the $k_y$ direction. In the approach proposed here, the resulting density of states (DOS) [Fig. 2(d)] has a clear V-shape feature and vanishes at the energy where the two new Dirac cones meet, well separated from other states [36]. This is an important feature for ease in probing and utilizing the emergent Dirac fermion states [4,37]. We have examined two other potentials, corresponding to those of molecule assembly and patterned gate (with achievable parameters) in experiments [36]. Generally, if the potential deviates from Eq. (3), the above procedure remains valid. But the generated Dirac cone may be slightly shifted away from $K_1$ [12] and have other characteristics as mentioned above. Numerical simulations show that the other two potential models also produce sizable massless Dirac cones in an isolated energy window, and can be characterized in experiments such as scanning tunneling microscope and electrical transport, etc. [36]. The above procedure thus gives a general scheme for generating massless Dirac fermions in an arbitrarily anisotropic 2DEG from the perspective of inverse design.

Next, we explore one of the most counterintuitive phenomena in graphene, Klein tunneling. That is, normal-incident carriers in graphene experience a perfect transmission through a potential barrier independent of the potential width and height, made possible by the chiral nature of the linearly dispersing Dirac fermions and charge-conjugation symmetry [30]. In an anisotropic Dirac cone, for states with wavevector $k$ (measured from the Dirac point), the wavevector $k$, the pseudospinor $s$ and the group velocity $v$ are generally noncollinear [4]. In fact, they are given [for the potential in Eq. (3)] by $k = (k_x, k_y)$, $s = \lambda \hbar (k_x, \gamma_0 k_y)/\sqrt{k_x^2 + \gamma_0^2 k_y^2}$, and $v = \frac{\lambda v_0}{2}(k_x, \gamma_0^2 k_y)/\sqrt{k_x^2 + \gamma_0^2 k_y^2}$ and related by [Fig. 3(a)]

$$\tan \phi_v = \gamma_0 \tan \phi_s = \gamma_0^2 \tan \phi_k = \gamma_0^2 \frac{k_y}{k_x}, \tag{7}$$

where $\phi_k$, $\phi_s$ and $\phi_v$ are referenced to $k_x$ axis (note that $s$ depends on $\lambda = \pm 1$ but $\phi_s$ does not, by definition). If the relevant Fourier components of a potential deviate from Eq. (3) slightly, Eq. (7) is then an approximation.

Now we consider transmission through a potential barrier with a height of $V_0$ and width of $D_0$ ($D_0 \gg |a_1|, |a_2|$). Let us align the minor and major axes of the elliptical Fermi surface along the $k_x$ and $k_y$ direction, respectively (Fig. 3). We place a potential barrier in the $x'$-$y'$ coordinate system (rotated at some arbitrary angle $\alpha$ from the $x$-$y$ coordinate system), and make the barrier axis infinitely long along the $y'$ direction with the width $D_0$ lying along the $x'$ direction. In the continuum limit, translational symmetry along the $y'$ direction will conserve the $k_{y'}$ wavevector component of a propagating wave impinging on the potential barrier, based on which we look for the solutions in the three regions – before (I), inside (II) and after (III) the barrier [36]. The various notations are defined in the caption of Fig. 3.

The transmission probability in the ballistic transport limit is obtained by matching the carrier wavefunction at the boundaries [2,30]. The solutions of the three regions in the $x'$-$y'$ coordinate system take the form

$$\begin{cases} \psi_I(x',y') = \frac{1}{\sqrt{2}}\begin{pmatrix} 1 \\ \lambda e^{i\phi_s} \end{pmatrix} e^{i(k_{x'}x'+k_{y'}y')} + \frac{r}{\sqrt{2}}\begin{pmatrix} 1 \\ \lambda e^{i\phi_s^r} \end{pmatrix} e^{i(k_{x'}^r x' + k_{y'}y')}, & x' < 0, \\ \psi_{II}(x',y') = \frac{a}{\sqrt{2}}\begin{pmatrix} 1 \\ \lambda' e^{i\theta_s} \end{pmatrix} e^{i(q_{x'}x'+k_{y'}y')} + \frac{b}{\sqrt{2}}\begin{pmatrix} 1 \\ \lambda' e^{i\theta_s^r} \end{pmatrix} e^{i(q_{x'}^r x' + k_{y'}y')}, & 0 < x' < D_0, \\ \psi_{III}(x',y') = \frac{t}{\sqrt{2}}\begin{pmatrix} 1 \\ \lambda e^{i\phi_s} \end{pmatrix} e^{i(k_{x'}x'+k_{y'}y')}, & x' > D_0, \end{cases} \tag{8}$$

The transmission amplitude is

$$t(\phi'_v) = \frac{\lambda\lambda' e^{-ik_{x'}D}\left(e^{i\theta^r_s} - e^{i\theta_s}\right)\left(e^{i\phi_s} - e^{i\phi^r_s}\right)}{A} \qquad (9)$$

where

$$\begin{aligned}A = &\ e^{-iq_{x'}D}\left(e^{i\theta_s+i\theta^r_s} + e^{i\phi_s+i\phi^r_s} - \lambda\lambda' e^{i\theta^r_s+i\phi^r_s} - \lambda\lambda' e^{i\theta_s+i\phi_s}\right) \\ &- e^{-iq^r_{x'}D}\left(e^{i\theta_s+i\theta^r_s} + e^{i\phi_s+i\phi^r_s} - \lambda\lambda' e^{i\theta^r_s+i\phi_s} - \lambda\lambda' e^{i\theta_s+i\phi^r_s}\right).\end{aligned} \qquad (10)$$

The transmission probability through the potential barrier is $T(\phi'_v) = tt^*$. Besides the resonant unit transmission at specific angles owing to a Fabry-Pérot like effect, the pseudospin-momentum locking [Eqs. (6)–(7)] gives rise to a complete suppression of the scattering only under the condition $\mathbf{k}' \to -\mathbf{k}'$. Considering the fact that $k_{y'}$ is conserved, this process will happen only when $k_{y'} = 0$. Consequently, $\phi^r_s = \pi + \phi_s$, $\theta^r_s = \pi + \theta_s$, $\phi_s = \theta_s$ and $q^r_{x'} = -q_{x'}$, resulting in $T = 1$, i.e. perfect transmission independent of the potential barrier height and width.

With an anisotropic Dirac cone, perfect transmission therefore occurs when the incident wavevector $\mathbf{k}$ is along the normal to the potential barrier. However, the fact that the group velocity and the wavevector are generally noncollinear leads to a remarkably distinct Klein tunneling behavior compared with graphene. For a wave packet, it is the group velocity, not wavevector, that describes the direction of center-of-mass motion and energy flow. In graphene, normal incident wave packet (or energy flow) is unimpeded; however, in the anisotropic case, the transmission probability of a normal incident wave packet is not unity and can be tuned and controlled. Furthermore, the normal incident direction and the perfect transmission direction are in general different, and the difference depends on $\alpha$ and is maximized at $\alpha_m = \arctan\frac{1}{\gamma_0}$, taking on a value of $\phi'_{v,m} = $ –50.8° with the effective masses considered here for black phosphorus [36]. The directional dependence of the transmission probability $T(\phi'_v)$ is plotted in Fig. 4, with varying incident angle $\phi'_v$ of the group velocity relative to the potential barrier normal, and in principle is measurable from a directionally dependent nonlocal resistivity experiment [38].

Because of the asymmetric Klein tunneling behavior, the transmission probability of normal incident wave packets can be controlled with different barrier alignment, height and width, whereas as discussed above the perfect transmission direction depends on $\alpha$ only. In the cases where the Fermi surface is smaller inside the potential barrier region than the outside regions, then at certain angles, no propagating wave solutions exist. But evanescent wave solutions are allowed by letting $q_{x'} = i\kappa'$ and $q^r_{x'} = i\kappa'_r$ ($\kappa', \kappa'_r \in \mathbb{R}$). Therefore, a beam of normal-incident wave packets can be completely turned off by tuning $V_0$, which is also a missing feature in pristine graphene or any isotropic Dirac fermion system.

The phenomena found in this work are general and applicable to any anisotropic host systems. Moreover, the emergence of highly tunable and easily accessible anisotropic massless Dirac fermions in few-layer black phosphorus superlattices should provide a range of interesting experimental investigations and a new direction for possible device applications.

**Acknowledgements**

The authors thank L. Li, A. Rubio, F. H. da Jornada and G. Antonius for fruitful discussions. This work was supported by the Theory of Materials Program at the Lawrence Berkeley National Laboratory through the Office of Basic Energy Sciences, U.S. Department of Energy under Contract No. DE-AC02-05CH11231 which provided theoretical analyses and calculations of generated anisotropic Dirac fermions; and by the National Science Foundation under Grant No. DMR-1508412 which provided numerical simulations of Klein tunneling. Computational resources have been provided by the National Energy Research Scientific Computing Center, which is supported by the Office of Science of the U.S. Department of Energy, and the Extreme Science and Engineering Discovery Environment (XSEDE), which is supported by National Science Foundation Grant No. ACI-1053575.

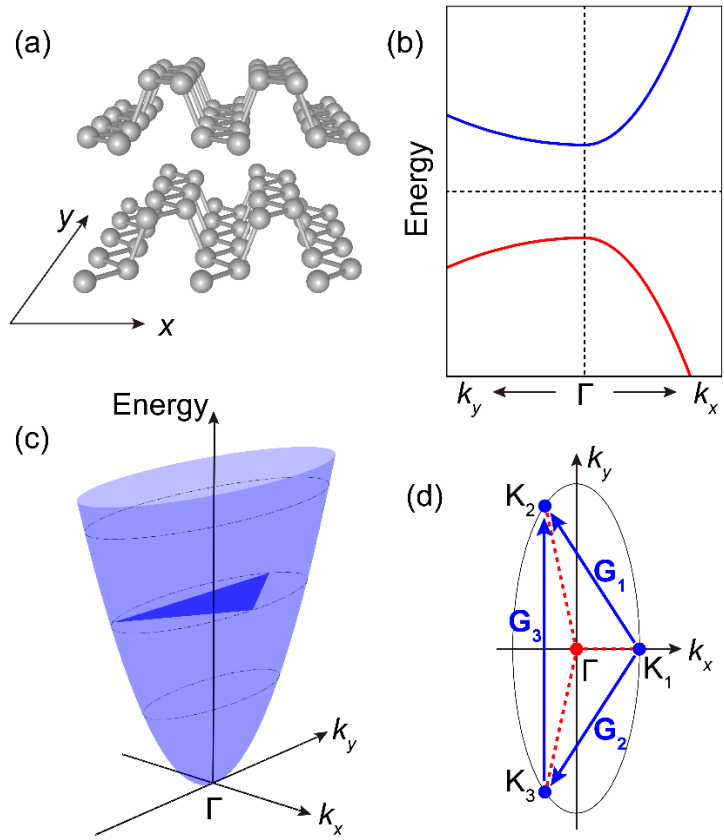

FIG 1. Few-layer black phosphorus. (a) Crystal structure and orientation. (b) Schematic band structure near the band gap around the Γ point. Both the lowest conduction band (blue curve) and the highest valence band (red curve) disperse much faster along the $k_x$ direction (armchair direction) than the $k_y$ direction (zigzag direction). (c) Geometry of a triangle defined by three points with degenerate energy in the reciprocal space, fitting into the band dispersion. (d) Elliptical isoenergetic contour. Three $\boldsymbol{K}_j$ points of degenerate energies on the energy surface define three $\boldsymbol{G}_j$ vectors. The reciprocal space of the superlattice is spanned by $\boldsymbol{G}_1$ and $\boldsymbol{G}_2$.

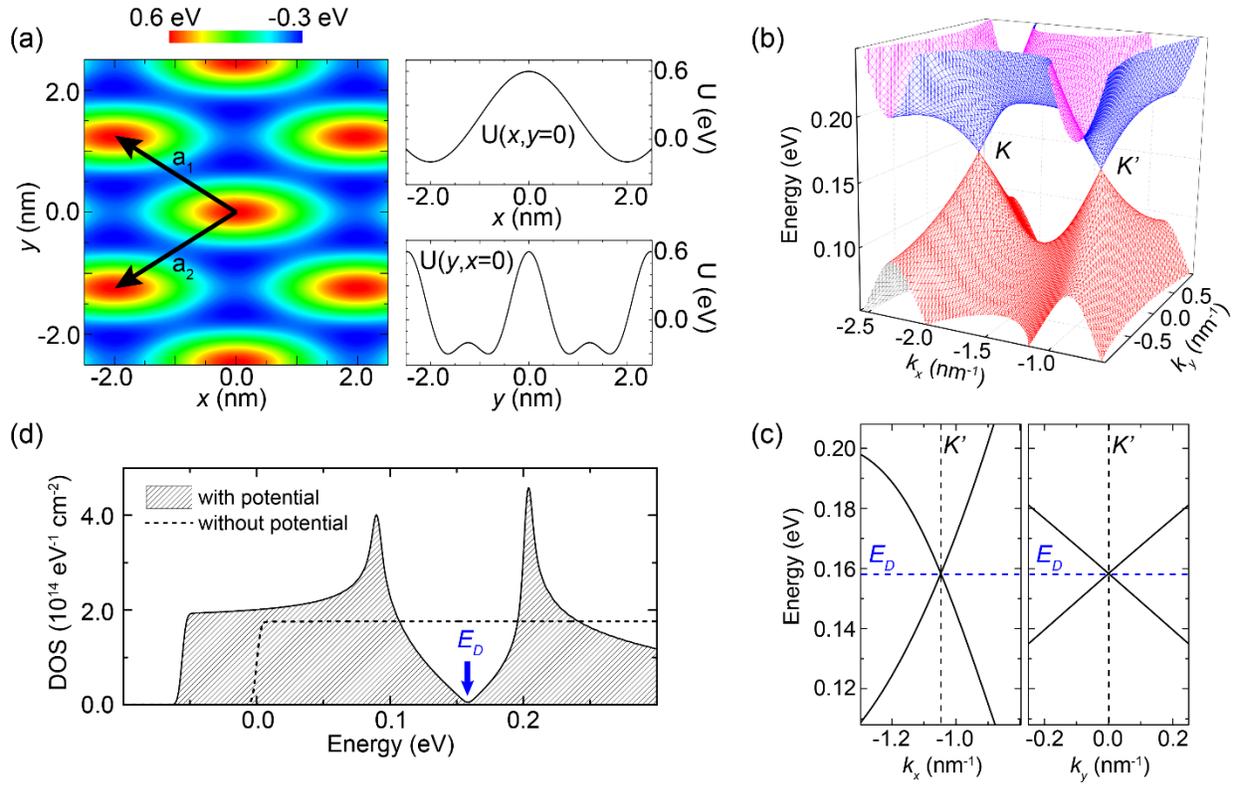

FIG 2. (a) Real-space distribution of a sinusoidal potential defined in Eq. (3) with $W = 0.1$ eV. $\boldsymbol{a}_1$ and $\boldsymbol{a}_2$ are the superlattice lattice vectors of the external periodic potential. $|a_1| = |a_2| = 2.35$ nm and the angle between the two vectors is 63.4 °. The two panels on the right are the line profiles of the external potential. (b) Band structures of the states under the sinusoidal periodic potential from non-perturbative numerical calculations. Anisotropic massless Dirac fermions are generated. (c) Band structures along two directions passing through one Dirac point, showing linearly dispersing features with strongly anisotropic group velocities. (d) DOS of the new Dirac fermions system generated with the external periodic potential. Dashed line represents the DOS from the lowest-energy conduction band in pristine few-layer black phosphorus before applying the external potential.

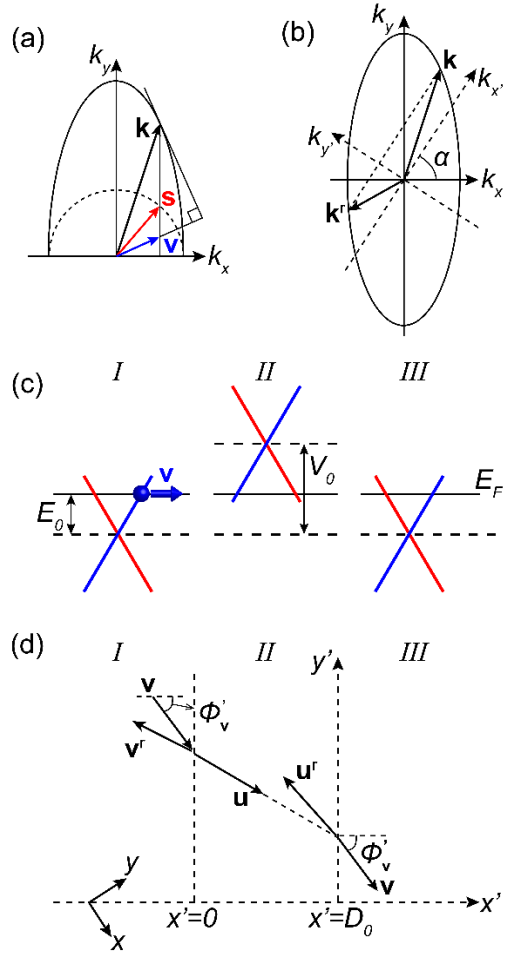

FIG 3. (a) Isoenergetic contour of the anisotropic Dirac cone. Wavevector $\boldsymbol{k}$, pseudospinor $\boldsymbol{s}$ and group velocity $\boldsymbol{v}$ are noncollinear in general. (b) With the elliptical Fermi surface plotted in the *x-y* coordinate system, a potential barrier is created with barrier normal along the x'-direction in the *x'-y'* coordinate system. Vectors and angles are denoted by the prime symbol if in the *x'-y'* coordinate system, and by $r$ if representing reflected waves. (c) An n-p-n junction (relative to the Dirac point) created by a potential barrier of height $V_0$ in region II. (d) Tunneling process through the potential barrier. $\boldsymbol{u}$, $\boldsymbol{u}^r$, $\boldsymbol{q}$, $\boldsymbol{q}^r$, $\theta$, $\theta^r$, $\lambda'$ in region II correspond to $\boldsymbol{v}$, $\boldsymbol{v}^r$, $\boldsymbol{k}$, $\boldsymbol{k}^r$, $\phi$, $\phi^r$, $\lambda$ in region I.

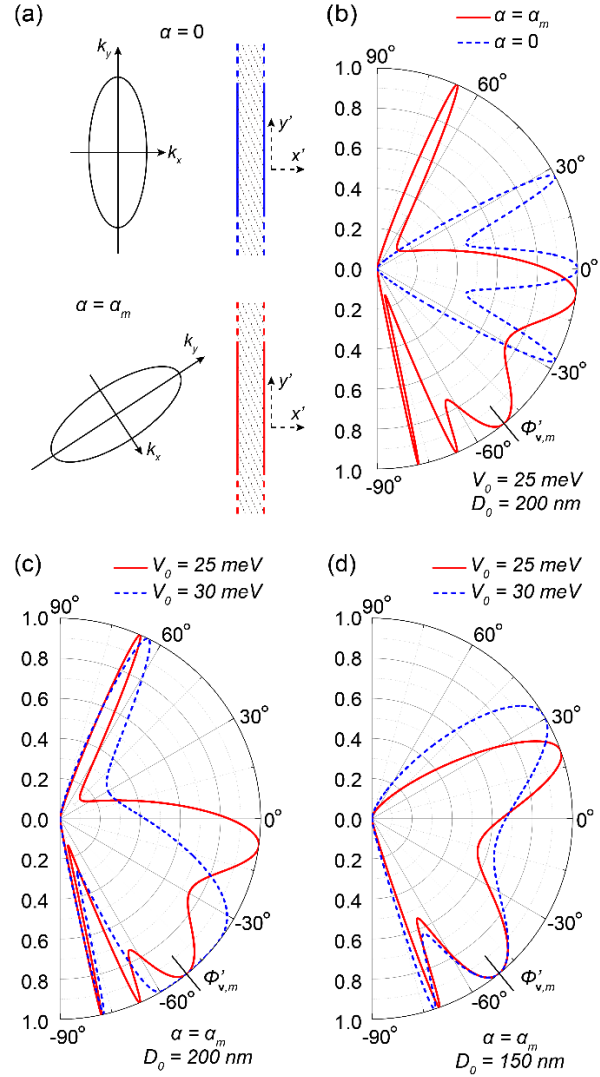

FIG 4. Transmission probability ($T$) versus the incident angle of group velocity ($\phi'_v$) with respect to the potential barrier normal, showing asymmetric Klein tunneling. $E_0 = 10$ meV is used. (a) Two different potential barrier orientations. At $\alpha = \alpha_m = \arctan\frac{1}{\gamma_0}$, the normal-incidence direction and the perfect-transmission direction are maximally differed by $\phi'_{v,m} = -50.8°$. (b) Symmetric and asymmetric Klein tunneling profiles corresponding to the two geometries in (a). (c)–(d) Asymmetric Klein tunneling with various parameters for $\alpha = \alpha_m$.

Supplemental Material for

# Generation of Anisotropic Massless Dirac Fermions and Asymmetric Klein Tunneling in Few-Layer Black Phosphorus Superlattices

Zhenglu Li, Ting Cao, Meng Wu, and Steven G. Louie[*]

*Department of Physics, University of California at Berkeley, Berkeley, California 94720, USA and Materials Sciences Division, Lawrence Berkeley National Laboratory, Berkeley, California 94720, USA*

[*]E-mail: sglouie@berkeley.edu

## I. Two realistic potential profiles

In the main text, our initial discussions are based on a simple form of potential, as defined in Eq. (3). Here we give details of two realistic cases also considered – a conical potential mimicking the molecular assembly [Fig. S1(a)], and a rectangular potential which can be realized with patterned electrostatic gating [Fig. S2(a)]. A potential can in general be written as the superposition of its Fourier components, i.e., $U(\mathbf{r}) = \sum_{\mathbf{G}} U(\mathbf{G}) e^{i\mathbf{G}\cdot\mathbf{r}}$. Since the potential is inversely designed from Eqs. (2)–(5) in the main text, the lowest $\mathbf{G}$ components, which defines the size and shape of the superlattice, are similar to those defined in Eq. (3) in the main text. In the conical potential, a small deviation from Eq. (3) exists and effectively shifts the positions of the Dirac points away from the $\mathbf{K}_j$ points. However, if the profile of the potential is very localized (such as in molecule assembly), the analysis in the main text can still be treated as a good approximation to the realistic situation. Therefore, our conclusions on the generation of massless Dirac fermions around $\mathbf{K}_j$ points are valid, despite the fact that the detailed band structure in other regions away from $\mathbf{k} = 0$ may vary depending on the higher $\mathbf{G}$ components of the potential. The periodicity of the conical potential is 2.35 nm, which is quite accessible in experiments, *e.g.*, the assembly of monoxide molecules on Cu (111) surface takes a periodicity of less than 2 nm. The periodicity of the rectangular potential is 7.84 nm, in the same order of magnitude of what has been achieved with current lithographic techniques that can usually reach 10 – 20 nm.

Fig. S1(b)–(c) and Fig. S2(b)–(c) show the band structures from the two types of potentials, respectively. The anisotropic linear dispersions are clearly seen. Note that for both potentials, the DOS shows the V-shape feature and a vanishing point.

## II. Isolation of the new Dirac fermions within an energy window

Applying the external potential shifts and redistributes the DOS, creating an energy windows where only those newly generated anisotropic Dirac fermions exist. In contrast, tunable and anisotropic Dirac fermions can also be generated in graphene using one-dimensional periodic potential due to the chiral nature of states in the original Dirac cone, however, are typically obscured by other states in the same energy window (as discussed in

Ref. 4 and Ref. 37). In the approach proposed in this work, the resulting new states are well separated from other states. Thus, they can be more easily characterized in many experiments such as scanning tunneling microscope, electrical transport, quantum oscillations, quantum Hall, magnetoresistance and electron optics (e.g., different behaviors in phenomena such as electron focusing, negative refraction, anomalous reflection, *etc.*). Comparison with the DOS of the pristine few-layer black phosphorus [Fig. 2(d) in the main text] indicates that the system is only slightly modified by the superlattice potential, and the intrinsic band gap of few-layer black phosphorus (~ 1 eV) is still well preserved. Therefore, in such a black phosphorus superlattice device, the semiconducting phase and the massless Dirac fermion phase can be reversibly switched by tuning the Fermi level within an achievable carrier density range in the order of $10^{12} - 10^{13}$ cm$^{-2}$.

Noticing that the DOS plotted in Fig. 2(d), Fig. S1(d), and Fig. S2(d) only contain the effects of the external potential on the original lowest-energy conduction band, whereas in a real material, other bands need to be considered as well, because they may appear in the same energy window where the Dirac fermions exist, and therefore affect the experimental measurements. Previous *ab initio* calculations (Ref. 27 in the main text) show that in a monolayer black phosphorus, there are other bands ~ 0.2 eV higher in energy than the conduction band minimum and ~ 0.5 eV in energy lower than the valence band maximum. When the number of layers increases to four, other conduction band states almost reach the same energy as conduction band minimum, but those in the valence bands are still at ~ 0.5 eV away from the valence band maximum. Considering the energy of the new Dirac points and potentially interfering states, hole doping in few-layer black phosphorus should be more robust in the generation of isolated-in-energy anisotropic Dirac fermions. However, the ratio of the effective masses along two crystal axes in the hole doping regime varies a lot with number of layers ($\gamma_0$ is ~ 1/6 in monolayer with hole doping), while it is almost a constant as a function of layer number in conduction bands. Therefore, much care is needed in designing the external potential for a sample with a particular number of layers and the choice of electron or hole doping. Nevertheless, our analyses and calculations in the main text and above based on the conduction band states are quite general, and can be easily applied to the valence band and other systems having anisotropic two-dimensional electron gases.

### III. Solving for asymmetric Klein tunneling

The crystal axes of few-layer black phosphorus and the superlattices are defined in the *x-y* coordinate system, and the potential barrier is defined in the *x'-y'* coordinate system. Vectors and angles in the *x'-y'* coordinate system are denoted with a prime symbol. We consider group velocity $\boldsymbol{v}$ of the incident wave packet that can hit the potential barrier, so $\phi'_v \in \left[-\frac{\pi}{2}, \frac{\pi}{2}\right]$. With a given $\phi'_v$ ($\phi_v = \phi'_v + \alpha$), because $\tan\phi_{\boldsymbol{k}} = \frac{k_y}{k_x}$ and $\tan\phi_v = \gamma_0^2 \frac{k_y}{k_x}$, $\phi_{\boldsymbol{k}}$ can be determined (depending on $\lambda = \pm 1$), and we can write the incident wavevector in region I as $\boldsymbol{k} = (k_x, k_y) = (|k|\cos\phi_{\boldsymbol{k}}, |k|\sin\phi_{\boldsymbol{k}})$, satisfying

$$E_0 = \lambda \frac{\hbar}{2} v_0 \sqrt{k_x^2 + \gamma_0^2 k_y^2} = \lambda \frac{\hbar}{2} v_0 \sqrt{k^2(\cos^2\phi_{\boldsymbol{k}} + \gamma_0^2 \sin^2\phi_{\boldsymbol{k}})}, \tag{S1}$$

and we then have

$$|k| = \frac{2|E_0|}{\hbar v_0}(\cos^2\phi_{\boldsymbol{k}} + \gamma_0^2 \sin^2\phi_{\boldsymbol{k}})^{-\frac{1}{2}}, \tag{S2}$$

The same vector is represented in the $x'$-$y'$ coordinate system as $\boldsymbol{k} = (|k|\cos\phi'_k, |k|\sin\phi'_k)$ where $\phi'_k = \phi_k - \alpha$. In region I, the reflected wavevector takes $\boldsymbol{k}^r = (k^r_{x'}, k_{y'})$ in the $x'$-$y'$ coordinate system, because of $k_{y'}$ conservation. In the $x$-$y$ coordinate system, the reflected wavevector reads

$$k^r_x = k^r_{x'}\cos\alpha - k_{y'}\sin\alpha,$$
$$k^r_y = k^r_{x'}\sin\alpha + k_{y'}\cos\alpha, \tag{S3}$$

satisfying

$$k^{r2}_x + \gamma_0^2 k^{r2}_y = \frac{4E_0^2}{\hbar^2 v_0^2}, \tag{S4}$$

and we get the solution of $k^r_{x'}$, as well as the spinor angle $\phi^r_s$. In region III, there is only one wavevector which is the same as $\boldsymbol{k}$.

In region II, the wavevectors are denoted as $\boldsymbol{q}$, $\boldsymbol{q}^r$, with $q_{y'} = q^r_{y'} = k_{y'}$, satisfying

$$E_0 = \lambda' \frac{\hbar}{2} v_0 \sqrt{q_x^2 + \gamma_0^2 q_y^2} + V_0. \tag{S5}$$

where

$$q_x = q_{x'}\cos\alpha - k_{y'}\sin\alpha,$$
$$q_y = q_{x'}\sin\alpha + k_{y'}\cos\alpha, \tag{S6}$$

and equivalently, by solving

$$q_x^2 + \gamma_0^2 q_y^2 = \frac{4(V_0 - E_0)^2}{\hbar^2 v_0^2}, \tag{S7}$$

we can have two solutions for $q_{x'}$, one for incident and the other for reflected wave in region II, depending on the group velocities they associated with. Therefore we obtain the spinor angles $\theta_s$ and $\theta^r_s$. Note that Eq. (6) and Eq. (7) in the main text should be combined (*i.e.* considering $\lambda(\lambda') = \pm 1$) to determine which quadrant the wavevector and the group velocity lies in, and the spinor direction is determined by spinor angle and $\lambda$ ($\lambda'$) together. In a word, to obtain the solutions, the defining equations of the three vectors should be used, coming from Eq. (6) in the main text.

In a general asymmetric Klein tunneling process, the normal incidence direction is different from the perfect transmission direction, and this difference can be maximized with given $m^*_x$ and $m^*_y$. The perfect transmission case corresponds to $\boldsymbol{k} = (k\cos\alpha, k\sin\alpha)$, according to the main text. In this case, if we limit $0 < \alpha < \frac{\pi}{2}$, we can maximize $(\alpha - \phi_v)$, or equivalently $\tan(\alpha - \phi_v)$. Together with Eq. (7) in the main text, we may solve for the maximum value (at $\alpha = \alpha_m$) of the following function

$$\tan(\alpha - \phi_v) = \frac{\tan\alpha - \tan\phi_v}{1 + \tan\alpha\tan\phi_v} = \frac{(1 - \gamma_0^2)\tan\alpha}{1 + \gamma_0^2\tan^2\alpha}. \tag{S8}$$

Finally, we have

$$\tan\alpha_m = \frac{1}{\gamma_0}. \tag{S9}$$

The maximum $(\alpha - \phi_v)$ corresponds to $\tan\phi_{v,m} = \gamma_0$, and meanwhile $\tan\phi_{s,m} = 1$. Therefore, $\phi'_{v,m} = \phi_{s,m} - \alpha = \arctan\gamma_0 - \arctan\frac{1}{\gamma_0}$.

With the alignment of the potential barrier and the solutions in the three regions determined, starting from Eq. (8), we can have the following set of equations from the boundary conditions,

$$\begin{cases} 1+r = a+b, \\ \lambda e^{i\phi_s} + \lambda r e^{i\phi_s^r} = \lambda' a e^{i\theta_s} + \lambda' b e^{i\theta_s^r}, \\ a e^{iq_{x'}D} + b e^{iq_{x'}^r D} = t e^{ik_{x'}D}, \\ \lambda' a e^{i\theta_s + iq_{x'}D} + \lambda' b e^{i\theta_s^r + iq_{x'}^r D} = \lambda t e^{i\phi_s + ik_{x'}D}. \end{cases} \qquad (S10)$$

With some algebra, Eqs. (9)–(10) in the main text can be reached.

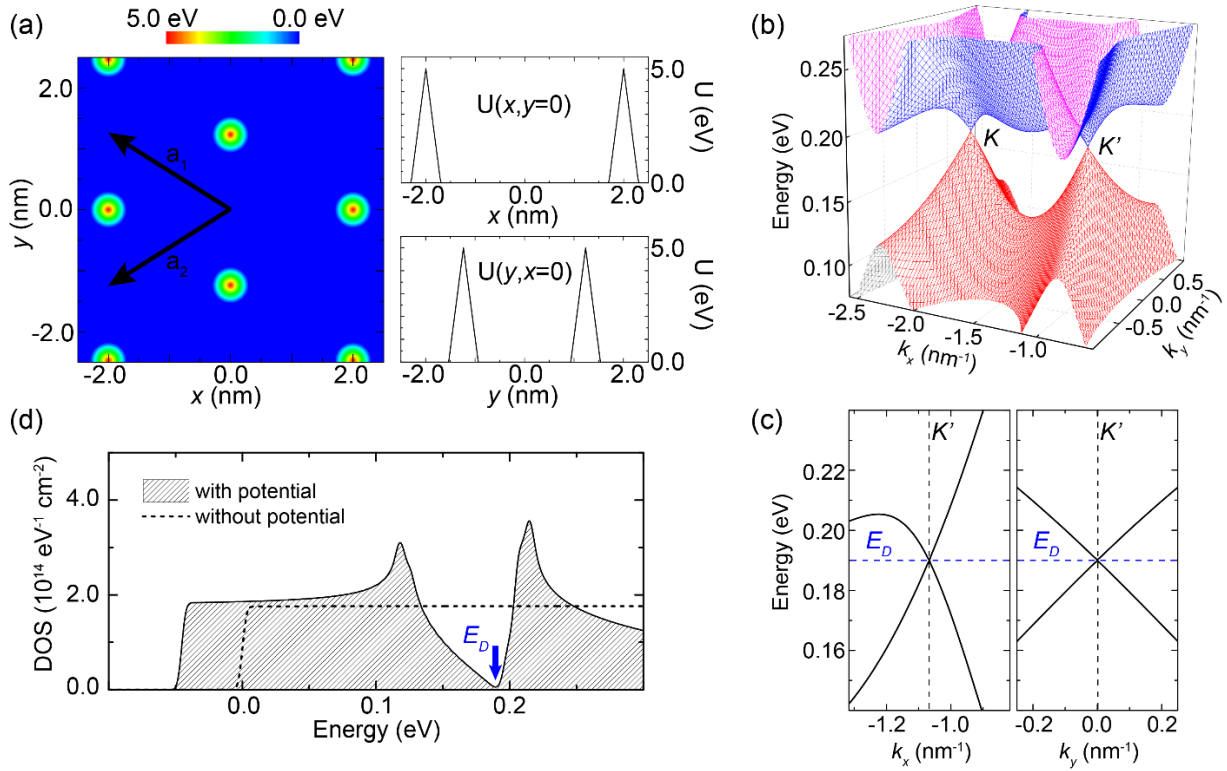

FIG. S1. Generation of massless Dirac fermions from a conical potential mimicking the molecule assembly. (a) Real-space distribution of the conical potential of radius 0.3 nm with 2.35 nm periodicity, along with two line profiles plotted in the right two panels. (b) Band structures. (c) Band structures along two normal directions passing through one Dirac point. (d) DOS showing V-shape feature with a vanishing point. Real electron spin degree of freedom is included.

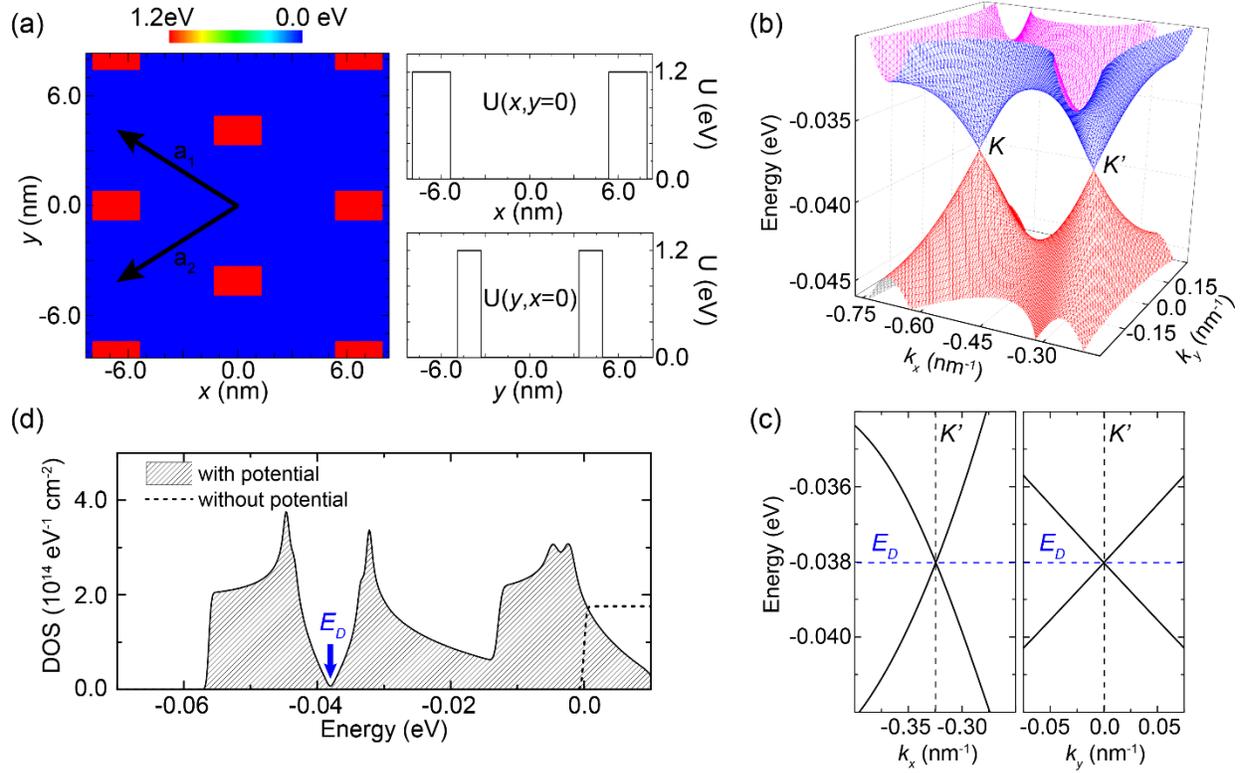

FIG. S2. Generation of massless Dirac fermions from a rectangular potential mimicking the patterned electrostatic gating. (a) Real-space distribution of the rectangular potential of 2.67 nm long and 1.65 nm wide with 7.84 nm periodicity, along with two line profiles plotted in the right two panels. Note that a different periodicity is used compared with the other two potentials, leading to a different energy scale. (b) Band structures. (c) Band structures along two normal directions passing through one Dirac point. (d) DOS showing V-shape feature with a vanishing point. Real electron spin degree of freedom is included.